# Three regimes in the tribo-oxidation of high purity copper at temperatures of up to 150°C

Julia S. Rau[a,b], Oliver Schmidt[a,b], Reinhard Schneider[c], Christian Greiner[a,b*]

[a] Karlsruhe Institute of Technology (KIT), Institute for Applied Materials (IAM), Kaiserstrasse 12, 76131 Karlsruhe, Germany. julia.rau@kit.edu; S-Oliver@gmx.de greiner@kit.edu

[b] KIT IAM-CMS MicroTribology Center µTC, Strasse am Forum 5, 76131 Karlsruhe, Germany.

[c] Karlsruhe Institute of Technology (KIT), Laboratory for Electron Microscopy (LEM), Engesserstrasse 7, 76131, Karlsruhe, Germany. Reinhard.schneider@kit.edu

*Corresponding author: greiner@kit.edu, +49 72120432742

**Grapical Abstract:**

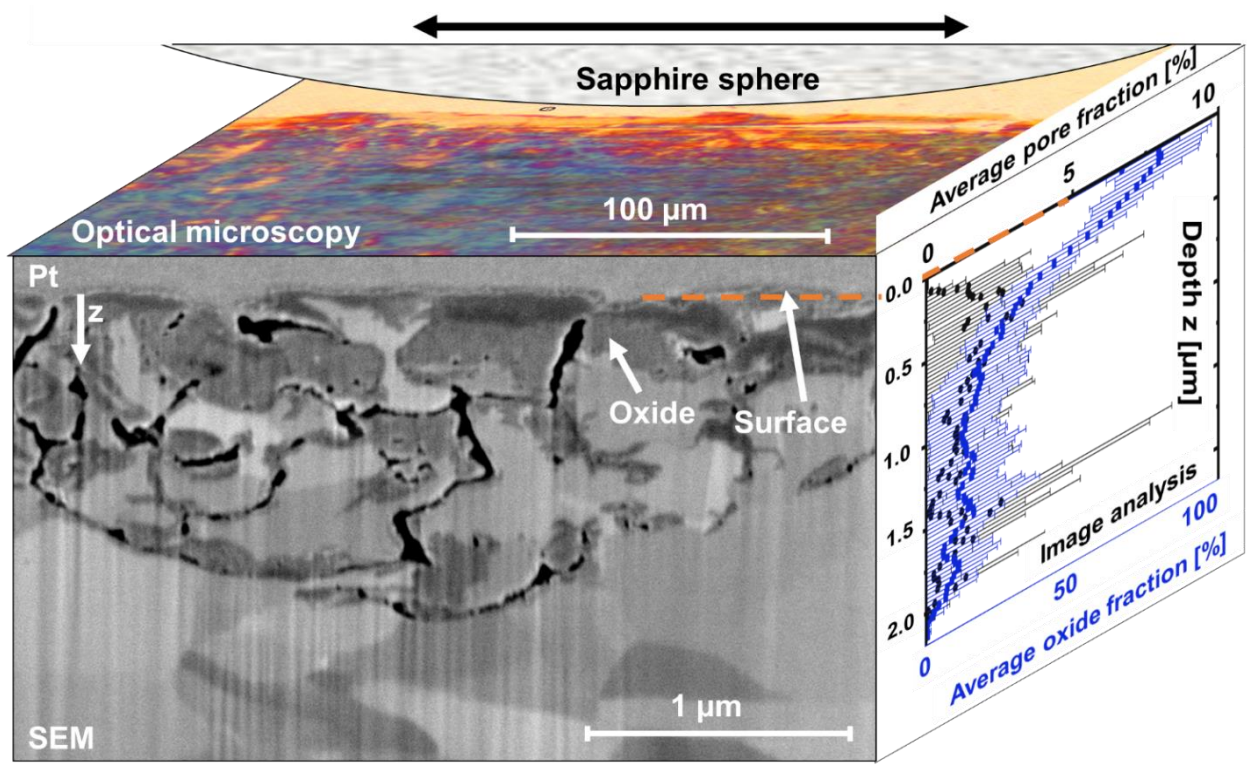

**Abstract:**

Surface oxidation of high-purity copper is accelerated under tribological loading. Tribo-oxide formation at room temperature is associated with diffusion processes along defects, such as dislocations or grain boundaries. Here, we embark on investigating the additional influence of temperature on the tribo-oxidation of copper. Dry, reciprocating sliding tests were performed with a variation of the sample temperature between 21 – 150 °C. Microstructural changes were monitored and analyzed with state-of-the-art electron microscopy techniques. Oxide layer formation through thermal oxidation was observed for 150 °C, but not for lower temperatures. As the temperature increases from room temperature up to 100 °C, a significantly stronger tribo-oxidation into deeper material layers and an increase in the amount of formed pores and oxides was detected. Up to 75 °C, diffusional processes along grain boundaries and dislocation pipes were identified. Starting at 100 °C, CuO was detected. Hence, tribological loading significantly alters the CuO formation in comparison with static oxidation. Along with the CuO formation at temperatures ≥ 90 °C, the oxide layer thickness decreased while the friction coefficient increased. The observations broaden our understanding of the elementary mechanisms of tribo-oxidation in high-purity copper. Eventually, this will allow to systematically customize surfaces showing tribo-oxidation for specific tribological applications.

**Main Text:**

1. **Introduction**

Chemical reactions are known to accelerate or to even take completely different pathways under tribological loading compared to static conditions[1,2]. Often, such behavior is associated with normal and shear stresses as well as a temperature increase at local asperity contacts[2]. Oxidic products are frequently observed on tribological-loaded surfaces[3–5]. The exact formation mechanisms of these tribo-oxides remain however unclear. Even without a tribological load, oxidation of metals is complex[6,7]. Among the most intensively studied metals is copper, which was investigated for a broad range of parameters[6]. For the same environmental conditions, different oxidation behavior was reported[7]. Copper's oxidation is decisively determined by oxidation temperature and oxygen partial pressure, surface state and morphology, oxidation duration, substrate temperature and specimen microstructure [6–9]. Hence, a strategic investigation of oxidation requires a careful choice of parameters and materials.

The two most common copper oxides are purple $Cu_2O$ and black $CuO$. Often, a thin layer of $CuO$ is reported to form ontop $Cu_2O$ layers[6,10]. $Cu_2O$ formation is predominant below 180 °C for static oxidation[7,8,11]. With increasing temperatures, the growth rate of the oxide layer increases[6,7]. Applying a mechanical load onto a copper surface adds more complexity to the oxidation process. Oxide formation under tribological load is called tribo-oxidation. Tribo-oxides grow with different rates compared to native oxidation and they can exhibit other morphologies compared to under static conditions[12,13]. For example, when compared with native oxidation, tribological loading accelerated oxidation in high-purity copper paired with a sapphire sphere up to a factor of 470[14]. Under sliding conditions, semicircular copper oxides formed, comprised of nano-crystalline $Cu_2O$ areas embedded in an amorphous oxygen-rich matrix[3]. In contrast to native oxidation, tribo-oxide islands grew into the material and not on the substrate surface[3,15].

As tribo-oxidation is widely known in literature, it is somewhat surprising that, so far, the exact elementary mechanisms for oxidation are not yet fully understood. This so far incomplete picture does not allow to strategically alter or make use of tribologically-induced chemical processes. In addition, no strategic design of materials more resistant to tribo-oxidation is feasible. In a recent set of experiments, the oxide formation mechanisms in high-purity copper under mild tribological loading at room temperature were associated with diffusion processes [3,14]. Tribologically-induced high diffusivity pathways, i.e. defects like phase/grain boundaries and dislocations, were identified as key for understanding tribo-oxidation under mild conditions[3,14]. However, it remains unclear how an increased sample temperature affects tribo-oxidation: when sliding at higher temperatures, thermal, mechanical and chemical mechanisms may all be active. This results in complex formation processes for the oxidation products, which might be significantly different from those found for room temperature sliding. Literature reports temperature having a strong influence on chemical reactions and occurring wear mechanisms[16],[17]. Such reactions often lead to a change in, for example the mechanical properties or the surface's chemical composition[18–20]. A beneficial wear behavior was reported along with the formation of compacted 'glaze layers' at higher temperatures[5,21]. However, brittle oxide layers may spall-off the surface and cause a strong increase in abrasive wear[22,23]. The question therefore arises how tribooxidation does alter for higher sample temperatures in high-purity copper. Do the reaction rate, the dominant diffusion pathways or the friction and wear behavior change? We here address these questions by increasing the sample temperature in sliding experiments from room temperature up to 150 °C. We are particularly interested in the role of the diffusion pathways and possible changes in mechanisms with increasing temperature. In terms of applications, devices made of copper and its alloys are often exposed to such higher temperatures. For example, electrical conductors and connectors in the automotive sector or heat exchangers operate in a temperature range between room temperature and 150 °C[24],[25].

## 2. Experimental:

**Materials:** High purity oxygen-free high conductivity (OFHC) copper samples with a purity > 99.95% (Goodfellow, Friedberg, Germany, 25 x 6 x 5 mm³) and an average grain size of 30 – 40 µm were paired with sapphire spheres (diameter of 10 mm, Saphirwerk, Bruegg, Switzerland). Copper is a model material to investigate the formation of oxides in general[6] and sapphire is used because of its chemical inertness and high hardness.

**Sample preparation:** The sample preparation followed our established routine, described in [3,26]. The copper samples were electro-polished right before the tests to achieve a sample surface with minimal native oxidation and no prior plastic deformation.

**Tribological testing:** Our reciprocating linear tribometer setup is described in [3,26,27]. The experimental parameters were 1.5 N normal load, 12 mm sliding distance, 1,000 sliding cycles and 1.5 mm/s sliding speed. The copper sample was moved by a linear motor (M-403.2PD, Precision Translation Stage, Physik Instrumente, Karlsruhe, Germany). The tests were performed with a reciprocating motion in air at a relative humidity of 50 % without lubrication. The humidity during the tribological experiments was closely controlled. We systematically varied the sample temperature ranging from 21, 50, 75, 90, 100, 125 °C to 150 °C by use of an electrical heating plate with a precision of ± 2 K. For a careful temperature control, the temperature sensor was clamped between the holder and sample to maintain a constant temperature. Two experiments for each temperature were performed besides 90 °C and 150 °C. Here, only one experiment was performed each to access more temperatures within the observed three regimes (Section 4.2). Right after the tribological tests, the copper samples were stored in a dessicator with a pressure < 1 mbar, to prevent additional oxidation due to the ambient environment.

**Characterization techniques:** The sapphire spheres and copper plates were analyzed after the experiments using a Sensofar Plµ Neox confocal microscope (Sensofar, Barcelona, Spain). Wear track depth was measured in confocal microscopy images. Microstructural analysis was performed by use of a focused ion beam / scanning electron microscope (FIB/SEM, FEI Helios 650, ThermoFisher Scientific, Waltham, Massachusetts, USA). Wear track width was measured in top-view SE-SEM images in the middle of the wear track. Cross-sections and transmission electron microscopy (TEM) foils were prepared directly after the tribological tests. The cross-sections were prepared along the sliding direction and perpendicular to the sliding surface in the center of the wear tracks (see SI of [3]). Two platinum layers (one with the electron and one with the ion beam) were deposited protecting the sample surface from ion beam damage. The cross-sectional images gave insights into the depth of the plastically deformed layer and the formation of possible oxides and pores. The maximum oxide thickness was determined on 7 – 16 images for each cross-section and averaged for each sample.

The oxide and pore distribution over the subsurface depth z were evaluated with ImageJ (NIH, Bethesda, Maryland, USA, Figure S4). The cropped region of interest which included the surface and the formed oxides was sliced in 20 nm thin sections and converted to binary black and white images. The threshold for this conversion was chosen such that pores or the oxides were represented as black areas. These black areas were measured with respect to the total sliced area. For each cross-section, four representative images were evaluated and averaged.

Energy-dispersive X-ray spectroscopy (EDXS) was performed inside a 200 kV TEM of the type FEI Tecnai Osiris equipped with a ChemiSTEM system with four silicon drift detectors (Bruker, Billerica, Massachusetts, USA) in the scanning TEM (STEM) mode for detailed microchemical analysis. Two TEM foils for two experiments with different temperatures were analyzed (100 and 125 °C). The minimum spot size of the electron probe during the EDXS measurements was smaller than 1 nm. The measurement period was 10 – 25 minutes, while a possible drift was automatically corrected by a cross-correlation with reference images.

Electron energy loss spectroscopy (EELS) of the same two samples was performed by means of a Gatan imaging filter Tridiem HR 865 (Gatan Inc., Pleasanton, USA) which is equipped to a TEM of the type FEI Titan³ 80-300 (Thermo Fisher Scientific, Waltham, USA). EEL spectra were recorded in the STEM mode with a dispersion of 0.5 eV per channel at 300 keV primary electron energy by means of a Gatan CCD camera UltraScan 1000 P. The size of the electron probe was approximately 0.5 nm and its convergence semi-angle 14 mrad. The energy resolution determined from the full width at half maximum (FWHM) of the zero-loss peak amounted to about 0.8 eV. The DigitalMicrograph software (Gatan) was used for spectrum recording and processing. A number of spectra were integrated with an individual measuring time of 1 s in the range from about 450 eV to 1450 eV to record the O-K edge (532 eV onset energy) as well as the Cu-$L_{23}$ edge at 931 eV. In addition, details of the energy loss near-edge structure (ELNES) of the O-K edge were studied by recording of spectra at 0.03 eV dispersion in the range between 525 eV and 585 eV.

3. Results:

**3.1 Tribological properties**

Within the first 100 cycles, the coefficient of friction (COF) increases, followed by a steady-state behavior (Figure 1). For temperatures up to 100 °C, the COF of the last 500 cycles is stable between 0.6 – 0.7. Since the COF for two experiments with the same temperature is almost identical, only one exemplary result is shown in Figure 1a. The experiments at 125 °C show smaller COFs between 0.4 – 0.6. For the 150 °C experiment, the highest values for the COF with ~ 0.8 - 1.0 are observed.

Figure S1 depicts representative optical microscopy images of the wear tracks and spheres (Figure S1a-g) and the average wear track width and depth with increasing temperature (Figure S1h). After 1,000 cycles, the sapphire spheres exhibit wear particles for all temperatures. The amount of wear particles increases with temperature. While lower temperature experiments exhibit only few particles (Figure S1a+b), temperatures above 50 °C show larger areas covered with wear particles. The 125 °C and 150 °C experiments further display scratches within the contact area (Figure S1f+g). The wear tracks on the copper plates all demonstrate a purplish color after 1,000 cycles. For 150 °C, a coloring of the 'unloaded' copper plate appears (Figure S1g). Starting at 90 °C, some dark/black areas inside the wear tracks are present. With increasing temperatures, the wear track broadens slightly from ~ 2.5 µm at 21 °C to 3.2 µm at 150 °C (Figure S1h). The depth of the wear tracks lies in the range of 1.2 – 1.7 µm for temperatures between 21 – 125 °C and increases slightly for the 150 °C experiment to ~ 2.25 µm (Figure S1h). The depth of the deformed layer, measured in cross-sectional SE-SEM images, increases for temperatures up to 75 °C to ~ 40 µm and remains stable between 35 – 40 µm for even higher temperatures (Figure S2).

### 3.2 Microstructural and chemical characterization

Figure 2 presents scanning electron microscopy images in a cross-sectional view after 1,000 sliding cycles with increasing experimental temperature. To protect the surface from ion beam damage, two platinum layers were applied; the surface is marked by a dashed line. After sliding at room temperature (Figure 2a), semicircular oxides and pores are observed. For sliding at 50 °C (Figure 2b), the oxides reach deeper into the bulk. The experiments at 75 °C (Figure 2c) show more pores, which are also connected horizontally. As for the 50 °C experiment, the oxide covers the whole surface. For higher sample temperatures (90 and 100 °C, Figure 2d+e) the oxides reach even further into the bulk. The pores form pathways and are closer to the surface. With temperatures below 90 °C, the pores are mainly present at the metal/oxide interface. For the 100 °C cross-section, oxidized areas are found adjacent to the pore pathways. The surface is less smooth than for the lower temperature experiments. At the surface, darker areas are present. Such darker areas are also found for the 125 °C experiment (Figure 2f+g). For the 150 °C experiment, the same behavior for the average oxide thickness like for the 125 °C experiment is observed (Figure 2h). For all temperatures, cross-sections next to the wear track were cut to investigate a possible thermal oxidation. Exemplarily, two cross-sections are shown in Fig. S3: up to 125 °C, no oxide layer outside the wear track can be resolved in the SE-SEM images. For the 150 °C experiment, a thin layer, containing pores, is present (Figure S3b). This sample further exhibits a color change next to the wear track from copper-colored to purple after 1,000 cycles (Figure S3c). The oxide thickness increases with temperature up to 100 °C (Figure 2j). From 21 to 75 °C the oxide thickness more than doubles (from ~ 0.21 µm to ~ 0.46 µm). Between 75 and 100 °C, the oxide thickness increases about factor 3 from ~ 0.5 to ~ 1.6 µm. Above 100 °C, it decreases again for the experiments at 125 °C and 150 °C.

The pore and oxide fraction of the SE-SEM images with increasing sample depth are measured by a quantitative image analysis method (Figure S4). Figure 3a+b summarize the average pore and oxide fraction for two exemplarly temperatures (21 and 150 °C, all other graphs are presented in Figure S6).

21, 50 and 75 °C all show a relatively similar behavior: within the first micrometer below the surface, the oxide fraction reaches zero. With increasing temperature, the pore fraction increases and pores are detected deeper inside the material. Starting at 90 °C, the oxide fraction displays two different slopes with increasing depth and in total, the oxides reach deeper into the material. The pore fractions show a wavy character. The image analysis provides additional information to analyze the resulting oxides: Figure 3c presents the total oxide and pore area as a function of sample temperature, summarized over all depths. Hence, no information about the oxide thickness is contained. Symbols without standard deviation represent single values. Up to 75 °C, the oxide area increases from 0.35 to 0.40 µm² and the pore area from 0.005 to 0.016 µm². Within the following 25 K temperature increase to 100 °C, the oxide area increases by about a factor 3 (from 0.4 µm² at 75 °C to 1.3 µm² at 100 °C) and the pore area by a factor 4 (from 0.016 µm² at 75 °C to 0.060 µm² at 100 °C). The oxide area stays almost constant for 125 °C sample temperature around a value of ~ 1.25 µm² in comparison with the 100 °C experiment. In-between 100 and 125 °C, there is a drop by roughly a factor of 6 for the total pore area (0.06 µm² at 100 °C to 0.01 µm² at 125 °C). For the 150 °C experiment, this value increases again to ~ 0.05 µm².

In Figure 2 dark areas in the near surface area are visible for T > 90 °C, which are identified as CuO (see below). Figure 3d depicts the CuO fraction for the temperatures between 100 and 150 °C. This is evaluated with the same method as the oxide fraction in Figure 3a+b. However, the grey value threshold is adjusted to only measure the dark oxide areas. The maximum CuO fraction has its maximum in-between the first 250 nm for all temperatures. The total CuO fraction increases with temperature (100 °C ~ 30 %, 125 °C ~ 55 % and 150 °C ~ 70 %).

Figure 4 schematically depicts the connection between the cross-sectional SE-SEM images in the middle of the wear track and the oxide- and pore-fraction analysis. To further investigate the nature of the dark oxides areas, energy-dispersive X-ray spectroscopy (EDXS) and electron energy loss spectroscopy (EELS) in the TEM for the 100 and 125 °C samples was performed (Figure 5). The EDS maps of the oxygen distribution in Figure 5a+d display areas with increased oxygen concentration in the near-surface zone. Inside the oxides, different oxygen concentrations are present. For 125 °C, an oxygen-rich 'arm' which reaches deeper into the oxide layer is visible in Figure 5e. Element concentration profiles for oxygen and copper along the white arrows in Figure 5b+e are depicted in Figure 5c+f. The lines represent the $Cu_2O$ and CuO composition. The experiment performed at 100 °C shows both, CuO composition at the surface and $Cu_2O$ composition closer to the bulk. For the 125 °C experiment, also two different oxygen compositions are present, however, the absolute amount of oxygen is lower than for the 100 °C sample. To gain more detailed insights into the nature of the oxides, a combined STEM/EELS analysis was performed on the same two samples (Figure 6). STEM brightfield and HAADF images of both samples are depicted in Figure 6a+c. The STEM-HAADF imaging mode allows to visualize local differences in the chemical composition because of its z-contrast sensitivity. Dark areas represent regions with locally lower density or smaller average atomic mass in comparison with copper. Figure 6b displays the ELNES-analysis of the O-K edge in the range between 525 and 585 eV: two different O-K edges in the near surface area and in the bulk oxide are present. Figure 6d shows the results of EELS point analyses of the O-K- and $Cu-L_{23}$-edges of the marked areas in Figure 6c for the 125 °C sample. The 125 °C experiment exhibits alternating layers of brighter and darker contrasts within the oxide (Figure 6c).

## 4. Discussion:

### 4.1 Effects of temperature on the tribological properties and wear track morphology

We first pose the question if temperature affects the tribological properties and resulting wear tracks and if yes in which way. Bowden and Tabor already pointed out that the COF of a system is to a large extent dependent on the yield stress of the softer material[28]. Metals' mechanical properties, such as the yield stress, may depend on temperature and usually decrease with increasing temperature[17]. We argue that the change in yield stress and Young's modulus of copper due to the increased temperature could not explain the overall COF behavior (see SI1).

Instead, a beneficial friction behavior of copper oxides in comparison with high-purity copper was previously reported[29]. Hence, along with the formation of copper oxides, lower COFs are expected. Here, temperatures up to 100 °C led to thicker oxides, while above 100 °C, the oxide thickness decreased, along with the appearance of dark oxide areas. At the same time, an almost constant coefficient of friction with increasing temperature (except for T > 100 °C) was measured (Figure 1a+b). Hence, the oxides' thickness does not alter the COF. Interestingly, the COF of the experiments above 100 °C seems to originate from the formation of the dark oxides, which here cover the surface (Figure 2 and Figure S1). TEM analysis revealed these areas to be CuO instead of $Cu_2O$. The nature and formation of these oxides will be discussed in Chapter 4.3. The COF of CuO was found to be ~ 0.1 higher (~ 0.27) than the COF of $Cu_2O$ (~ 0.18)[29]. This might explain the higher COF at 150 °C, with the highest CuO surface coverage ( Figure 3d), in comparison with the other experiments.

Figure S1 presents representative optical microscopy images of the wear tracks and spheres. Although previous results demonstrated no interaction with the sapphire sphere (for 0.5 mm/s, 12 mm sliding, 1,000 sliding cycles, 1.5 N normal load)[3], our findings show few wear particles on all counter bodies. With increasing temperature, more wear particles and scratches are visible on the spheres. Particles with similar appearance were previously found in the same type of contact and were characterized as comprised of copper, oxygen and aluminum[29]. We speculate about the influence of humidity in the contact in SI1.

**4.2 In search for diffusion pathways and mechanisms**

We pose the question: 'What do we learn from the temperature variation about the diffusion pathways and mechanisms responsible for oxidation?' First, the influence of temperature on the general oxidation properties is regarded. It is expected that temperature has an influence on the oxidation rate and thus oxide thickness. Next to the wear track, no oxide formation from thermal oxidation was observed in SE-SEM cross-sections up to 125 °C (Figure S3a). Only at 150 °C, the sample exhibited a color change and a few nanometer thin oxidation layer was detected (Figure S3b-d). However, when comparing the thermally grown oxide layer (few nm) with the tribological induced layer (~ 1 µm) for the 150 °C experiments, we can conclude that thermal oxidation plays only a minor role and all oxides within the wear tracks are mainly formed due to the tribological loading.

When comparing the oxide thickness with literature on native or thermal oxidation, the tremendous influence of the tribological loading becomes apparent: Up to 150 °C, literature reports a maximum oxide layer thickness of 30 - 100 nm[6,7] which is a factor 10 - 33 smaller than what we observed after sliding. Considering time as well, an oxide layer of ~ 3 nm was found on a thin copper film at 100 °C after roughly 4:40 hours, which corresponds to our test duration[30]. In comparison with the 1.6 µm thick layer in our experiments, tribological loading results in a factor ~ 530 thicker oxide.

The tremendous influence of tribological loading on oxide thickness is distinctive. Diffusion processes were found to be the key reason for this acceleration at room temperature[3],[14]. Do such processes also appear at higher temperatures? Important parameters for diffusion can be determined by the Arrhenius-equation for the diffusion coefficient $D(T)$:

$$D(T) = D_0 e^{-\frac{E_A}{RT}}$$

with the temperature independent diffusional constant $D_0$, activation energy $E_A$, universal gas constant $R$ and absolute temperature $T$. Fujita et al. found for low temperature oxidation (<180 °C) of copper single crystals that activation energies of oxidation agreed with the activation energies for diffusion (along surface and grain boundaries)[7]. They concluded that the low-temperature oxidation kinetics can be attributed for a large part to surface/grain boundary diffusion on/in the oxide. As a first approximation, we thus consider our results as predominantly diffusion controlled. We use the Einstein relation $D_O^{Cu} \sim \frac{d^2}{2t}$, to determine the diffusion coefficient of oxygen in copper for each temperature ($t$ = test duration, $d$ = oxide thickness)[31]. The maximum oxide thickness is used as the oxygen's distance covered in copper. The so calculated maximum diffusion coefficient represents an effective diffusion coefficient, taking into account all possible diffusion pathways (lattice, surface, grain- and phase boundaries as well as dislocation pipes). Such a simplified approach does not consider dynamic processes under sliding such as microstructural changes constantly creating and modifying diffusion pathways. It is to be viewed as a first order approximation, without taking changes in chemical composition or concentration gradients under loading into account. Since we measure the oxide thickness after 1,000 cycles, we do not distinguish between oxide formation and growth. Figure 7 shows the resulting Arrhenius plot. SI3 explains the larger standard deviation for the 125 °C experiments.

This Arrhenius analysis can be devided into three different and destinct regimes: from 21 – 75 °C (regime I), from 75 – 100 °C (regime II) and between 100 and 150 °C (region III). By fitting two linear fit functions for regimes I and II, the activation energies $E_A$ and diffusion constants $D_0$ are calculated: 15.4 kJ/mol in regime I and 106.5 kJ/mol in regime II. One might argue that considering the temperatures from 21 – 150 °C, all data points follow the same linear trend and the division in three regimes is an overinterpretation of the data. However, the 125 °C and 150 °C experiments were deliberately chosen as a third regime, as here the experiments showed a strongly deviating behavior. Moreover, due to the CuO formed in the wear tracks at temperatures above 90 °C, we expect a different mechanism here. We discuss this behavior in more detail later in 4.3.

Literature values for the activation energies for oxygen diffusion in $Cu_2O$ vary between 106 kJ/mol for diffusion along phase boundaries[32], 140 – 150 kJ/mol for lattice diffusion[33,34] and 160 kJ/mol along the surface[35] (measured at 800 – 1,120 °C). This broad range stems from the diverging measurements setups, parameters and measurement techniques used to determine diffusion coefficients of oxygen in copper. Very often, measurements are conducted at much higher temperatures (500 – 1,100 °C) than our experiments. Moreover, other sample dimensions, microstructures, oxygen partial pressures or thin films with different thicknesses are widely used[10,36,37]

As mentioned above, our values for the diffusion coefficients and activation energies represent effective ones, summarizing lattice, surface, grain boundary and dislocation pipe diffusion. The calculated activation energy for regime I with ~15 kJ/mol is surprisingly low; in comparison with the activation energy for diffusion along phase boundaries with 106 kJ/mol [32], a factor of 7 smaller. Hence, we assume that the low activation energy is a strong indication for the dramatic influence of the tribological loading, creating high diffusivity pathways such as grain boundaries or dislocations. For regime II, an activation energy of ~ 106 kJ/mol is calculated – a factor of 7 larger than for regime I. Processes within this temperature range exhibit a higher activation energy than for low temperature sliding. Hence, another or additional process most likely is activated. Usually, diffusion inside the more densely packed lattice requires higher activation energies compared to diffusion along defects, such as phase, grain boundaries and dislocations. For lattice diffusion of oxygen in $Cu_2O$ activation energies of 140 – 150 kJ/mol were reported[33,34], which is almost a factor 1.5 larger than what we observed. This might be indicative for the combined lattice and high diffusivity pathway active in this temperature regime, where higher temperatures provide more energy for lattice diffusion to be activated. Due to the simplyfied model applied for calculating the diffusion coefficients, the absolute values have to be interpreted with a grain of salt and are mainly used for comparison between regime I and II.

This analysis is solely based on the maximum oxide thickness. To gain more insights about the tribo-oxidation behavior with increasing temperature, a quantitative oxide and pore fraction analysis was performed (Figure 3). The accuracy of this analysis method is discussed in SI3. The quantitative oxide and pore analysis allows the following observations:

From the slope of the oxide fraction as function of subsurface depth, the oxide shape can be estimated. A large slope indicates flat oxides while a smale slope hints on the oxides reaching deeper into the material with a more semicircular shape. The total amount of pores and oxides increases stronger in-between 75 – 100 °C than for temperatures < 75 °C ( Figure 3c). The same behavior is observed in the Arrhenius analysis (Figure 7). A change in mechanism therefore is expected in these two regimes. The Arrhenius plot reveals different mechanisms for different temperatures; high diffusivity pathways up to 75 °C and an additional process above 75 °C (Figure 7). Earlier investigations found grain/phase boundaries and dislocation pipes as important pathways for diffusion under tribological loading at room temperature[14]. Do the profiles in Figure 3 reveal some of these pathways? For investigating the presence of grain boundary and pipe diffusion, the following analysis was performed:

Harrison established a widely used classification for diffusion kinetics in polycrystals, including three different types[38]. All types involve lattice diffusion from the surface, diffusion along a grain boundary or dislocation pipe as well as leakage into the bulk from the grain boundary/pipe in the vicinity around the high diffusivity pathway. Each type of pathway is prevailing for certain combinations of grain sizes, temperatures and diffusion times. In each regime, different diffusion profiles are expected. Our experiments were conducted on polycrystals; here the most widely observed type is B (Figure 8a). The grain boundary is described after Fisher[39]. For dislocation pipe diffusion, a model established by Smoluchowski is widely used[40]. A characteristic feature of type B is its penetration profile (Figure 8b): When plotting the logarithm of the concentration (c) of the diffusing species against the penetration depth $z^{6/5}$, type B has a profile that consists of two parts: direct lattice diffusion in the near-surface part (purple box) and a deeper straight 'grain-boundary tail' (blue arrow)[41]. Mehrer emphasized that the exponent of 6/5 has no reason grounded in physics, but is a good indicator for grain boundary diffusion[41]. Additionally, such a tail in the data can be considered as a measure for the quality of grain boundary diffusion detection in these analyses. Dislocation pipe diffusion leads to a linear behavior when log(c) is plotted against z[41]. This type of diagram allows to distinguish between diffusion along the lattice vs. along defects such as grain boundaries or dislocations. We approximated the concentration c with the oxide fraction representing the 'amount' of oxide at a certain depth z. Figure 8c-f displays the results of the oxide fraction logarithmically plotted against $z^{6/5}$ (indicative of grain boundary diffusion, Figure 8c+d) and z (indicative of dislocation pipe diffusion, Figure 8e+f) for average values.

Up to 75 °C, the profiles show a linear behavior for both $z^{6/5}$ and $z$ within the first 1,000 nm (Figure 8c+e). The slope for the first third of the penetration depth is almost the same; for the 50 and 75 °C experiments this is true even for two thirds of the depth. Such linear behavior (blue and red arrows) is indicative for grain boundary and dislocation pipe diffusion, however, distinguishing both in our experiments is very difficult[41]. For temperatures up to 75 °C (Figure 8c+e), no near-surface parts for lattice diffusion (purple boxes) are visible, indicating that at these lower temperatures, mainly high diffusivity pathways, induced by the tribological-load, contribute to oxide growth. This is in full agreement with the calculated activation energy of 15 kJ/mol. For temperatures above 75 °C (Figure 8d+f), the overall slope of the profiles is smaller. The profiles are less steady and show jumps (Figure 8d+f). Usually, such an analysis is performed when only one grain boundary or dislocation pipe is present. Since there are certainly more high-diffusivity pathways within the investigated cross-sections which are oriented differently with respect to the sample surface, the deviations from an ideal behavior is to be expected. With increasing temperatures (> 75 °C), a smaller slope in the near-surface area becomes more pronounced, which is associated with lattice diffusion (purple boxes, Figure 8b,d,f). This agrees with the calculation of the activation energy in regime II of (106 kJ/mol) (Figure 7). Temperatures above 100 °C are discussed in section 4.3.

Figure 8g presents a cross-sectional high-angle annular dark field (HAADF) STEM image after 1,000 cycles at 100 °C. Here, a pathway for fast diffusion along a grain boundary is visible. Additionaly, in the STEM-EDXS maps in Figure 5e an oxygen-rich pipe-like feature appears, connecting the surface with the bulk oxide. Collectively, these images and profiles confirm the different behavior observed in the Arrhenius plots (Figure 7): For temperatures up to 75 °C, mainly grain boundary and pipe diffusion are active. For higher temperatures, another process, possibly lattice diffusion, gets more and more important. Blau confirms that short-circuit pathways for oxidation become less important with increasing temperatures[17]. However, this was observed for a temperature increase above ~ 2/3 of the melting temperature (which would be ~ 630 °C for pure copper) and thus much higher than the maximum temperature studied here. In summary, by varying the temperature, we were able to identify different diffusion pathways dominating at different temperatures. Surprisingly, the oxide thickness decreases above 100 °C. We discuss this behavior in the following section.

**4.3 Tribological loading enhances CuO formation at temperatures ≥ 90 °C**

Regime III in Figure 7 was deliberately considered as an extra regime, since the experiments in this temperature range exhibited a behavior deviating from that for lower ones. One would expect deeper reaching oxides with increasing temperature as the mobility of the diffusing species is enhanced. However, Figure 2j reveals the oxides as less thick. A possible explanation could be that oxide wear particles form and spall-off the surface, thus reducing the measured oxide thickness. In contrast, no oxide wear particles formed on the copper plates for all temperatures and this explanation can be ruled out here (SI1). However, starting at sample temperatures of 100 °C, dark areas within the tribo-oxides evolve (Figure 2). By carefully taking into account the wear tracks presented in Figure S1, it is even possible to distinguish purple and dark areas with the naked eye. Interestingly, in Figure S1d, the experiment at 90 °C also showed some darker areas. Since they are not present in the middle of the wear track, they are not depicted in the SE-SEM cross-sections. STEM/EDXS and STEM/EELS reveal CuO formation for samples with temperatures of 100 and 125 °C (Figure 5 and Figure 6)[42]: Right below the surface, CuO was found, whereas $Cu_2O$ was observed closer to the copper bulk (Figure 5c+f). The lower oxygen composition of the 125 °C experiment originates from the larger thickness of this TEM-foil which resulted in absorption phenomena (Figure 5f). The layered structure with alternating $Cu_2O$ and CuO layers within the oxide after sliding at 125 °C indicates a succesive oxide growth by diffusion along the metal/oxide phase boundary (Figure 6).

From the analysis of Figure 8c-f and the activation energy calculated from Figure 7 one might speculate that the increased actvation energy in regime II results from the activation of lattice diffusion. Until now, only the diffusion of oxygen in copper oxide was investigated. However, copper is known to diffuse faster in copper oxide than oxygen (see also section 4.4). An increase in the lattice diffusion of copper would not lead to a larger oxygen contentration in the subsurface area but an increase in the copper concentration at the surface (Figure 5f). Which additional process needs to be considered here? A recent study found that tribo-oxidation under mild tribological loading might be rate-limited by the dissociation of oxygen at the oxide/air interface for mild sliding at room temperature[14]. Following this concept, dissociation rate constants for oxygen are known to increase with increasing temperatures[43]. Hence, higher temperatures not only lead to easier diffusion along the above described pathways, but also a higher availability of oxygen at the surface. Starting at 90 °C, this process provides enough oxygen to form CuO. CuO contains more oxygen and has a higher density compared to $Cu_2O$[44]. Ultimately, more oxygen is consumed directly at the surface to form CuO and the diffusion of oxygen into deeper areas is impeded. Consequently, the oxide thickness decreases in comparison to the lower temperatures, as more oxygen is needed to form CuO. Zhu et al. found that CuO scales are, in comparison to $Cu_2O$, very protective against oxidation, grow extremely slow and lattice diffusion is more difficult[44] (600 – 1,050 °C). The diffusion in CuO in this temperature regime is associated with the transport of atoms along high diffusivity pathways (such as dislocations and grain boundaries)[44]. The authors found that these fast pathways get blocked through a fresh oxide formation due to compressional stresses[44]. Ultimately, formed CuO layers reduce the diffusion for both, oxygen and copper within the oxide layer.

Astoundingly, in our experiments the formation of CuO takes place at temperatures significantly lower as expected from literature on static thermal oxidation[7]. Fujita et al. did not observe CuO until temperatures above 180 °C, while exposing a high purity copper single crystal to $Ar_2$ and $O_2$ for 24 hours[7]. O'Reilly et al. investigated oxidation of high purity copper up to 300 min in dry air between 50 and 500 °C[30]: Below 150 °C only $Cu_2O$ was observed, while $Cu_2O$ and predominantly CuO was only found for temperatures > 250°C. Native oxidation at room temperature sometimes yields a few nm thin layer of CuO on-top of the $Cu_2O$ layer[45]. However, Figure 3d clearly reveals CuO layer thicknesses of up to 250 nm for 100 °C, and a up to 500 – 1,000 nm thick CuO layer for 125 and 150 °C. Increasing temperatures increase the total amount of CuO at the surface ( Figure 3d).

Our explanation for this phenomenon is that tribological loading may change the energy landscape for CuO formation. Normal and shear stresses have been attributed to changes in the activation energies and reaction equilibria of chemical reactions[1]. Different reaction products under load than under static conditions were observed[2,12,13]. For example, tribopolymerization reactions of adsorbed molecules on surfaces through mechanical stresses were observed and the reaction pathway was different than for a chemical induced polymerization[2]. Another possible explanation is the pressence of water in the contact. The lack of water at higher contact temperatures due to evaporation, might indirectly lead to higher contact temperatures and with that earlier CuO formation, either through the absence of a lubricating film or a hardness change of the sapphire sphere[46]. As discussed in SI1, it remains unclear whether capillary bridges remain in the local asperity contacts at higher temperatures than 100 °C. In summary, the deviating behavior found for triologically-induced oxide formation in this temperature range compared to static oxidation is attributed to a tribologically-assisted formation of CuO, resulting in an overall lower oxide thicknesses.

**4.4 Pores and the Kirkendall-effect:**

In the previous chapters the oxide formation and its underlying processes were discussed in detail. Here we will instead highlight another feature within the tribologically-deformed region: pores. All SE-SEM images show pores, either within the oxide or at its boundaries (Figure 2). Pores in oxides under tribological loading are observed frequently[14,47,48] At temperatures above 21 °C, the oxides become less homogenous and the amount of pores increases (Figure 2, Figure 3a-c). Starting at 75 °C, the pores become more elongated (Figure 2), what hints to a faster diffusion along the pathways, possibly sub-grain boundaries, formed under tribological loading. Figure S6 shows the maximum pore area divided by the maximum oxide area plotted over the normalized oxide thickness. The maximum values are reached for deeper lying areas of the oxides (see Figure S8). Within the oxides, more pores form closer to the copper bulk than to the surface. This behavior can be associated with the Kirkendall-effect[49], which was previously observed also for room temperature experiments[14]. This effect has its roots in the diffusion of oxygen and copper within the oxides: For native oxidation, oxygen ions diffuse slower towards the copper-oxide interface than copper ions to the surface[50,51]. The faster outward diffusion of copper is compensated by an opposite flux of copper vacancies, creating the Kirkendall-pores at the metal-oxide interface[52]. Interestingly, the temperature does not influence the value of the normalized maximum oxide thickness where most pores are present, indicating that this effect is mainly controlled by the induced high diffusivity pathways, irrespective of temperature. Previous studies showed the accumulation of pores with increasing oxide thickness and eventually formation of cracks[14]. Such cracks can spall off material from the surface and ultimately lead to wear particles[14]. Although, the Kirkendall-effect was already observed at room temperature, the temperature variation here reveales the high-diffusivity pathways as principal underlying mechanism of this effect and as temperature independent.

## 5. Conclusion:

The effects of temperature on the tribo-oxidation of high-purity copper sliding against a sapphire sphere in a dry reciprocating tribological contact was systematically investigated. By use of state-of-the-art scanning and transmission electron microscopy and a thorough image analysis, the resulting microstructure was evaluated. The tribological properties and wear track morphology was investigated. In comparison with native or thermal oxidation, oxidation under tribological loading was accelerated by orders of magnitude for all temperatures. When increasing the sample temperature from 21 to 150 °C, these systematic investigations allowed to identify diverse diffusion pathways active at different temperatures and three seperate regimes were observed:

1. 21 – 75 °C: Tribo-oxide thickness increases with temperature. Low activation energy for diffusion (15 kJ/mol) and oxidation mainly governed by grain boundary and dislocation pipe diffusion.

2. 75 – 100 °C: Larger oxides than in regime I. Higher activation energy for diffusion (106kJ/mol). In addition to the processes active in regime I, lattice diffusion may become more important.

3. 100 – 150 °C: Reduction of oxide thickness in comparison to regime II. Formation of CuO at the sample surface and with that a sligth increase in COF. CuO observed at significantly lower temperatures than expected for native or thermal oxidation:

tribological loading eases CuO formation. No thermal oxidation was identifiable in SE-SEM images below 150 °C.

Kirkendall-pores were observed and found as being temperature-independent and resulting from diffusion along high-diffusivity pathways. These results allow to gain insights into the fundamental mechanisms of tribo-oxidation. This opens the door for a strategic tailoring of surfaces exhibiting tribo-oxidation and for enhancing the lifetime in many mechanical systems. More detailed work about the onset of tribo-oxidation in the early stages is required. Different sliding times will be needed to distinguish grain boundary and dislocation pipe diffusion in the diffusion profiles.

**Acknowledgments:**

Funding of this work has been provided by the European Research Council under ERC Grant Agreement No. 771237, TriboKey. We thank L. Pastewka for fruitful discussions.

**Data availability**
The data that support the findings in this study are available under the link https://doi.org/10.5445/IR/1000138897 and from the corresponding author upon request.

**Figure 1: Coefficient of friction (COF) with increasing temperature.** (a) COF as a function of increasing number of sliding cycles. All lines represent the average of two experiments, besides the experiment at 150 °C. (b) Average COF of the last 500 cycles. Each data point represents the average of two experiments, except for 150 °C where only one experiment was performed.

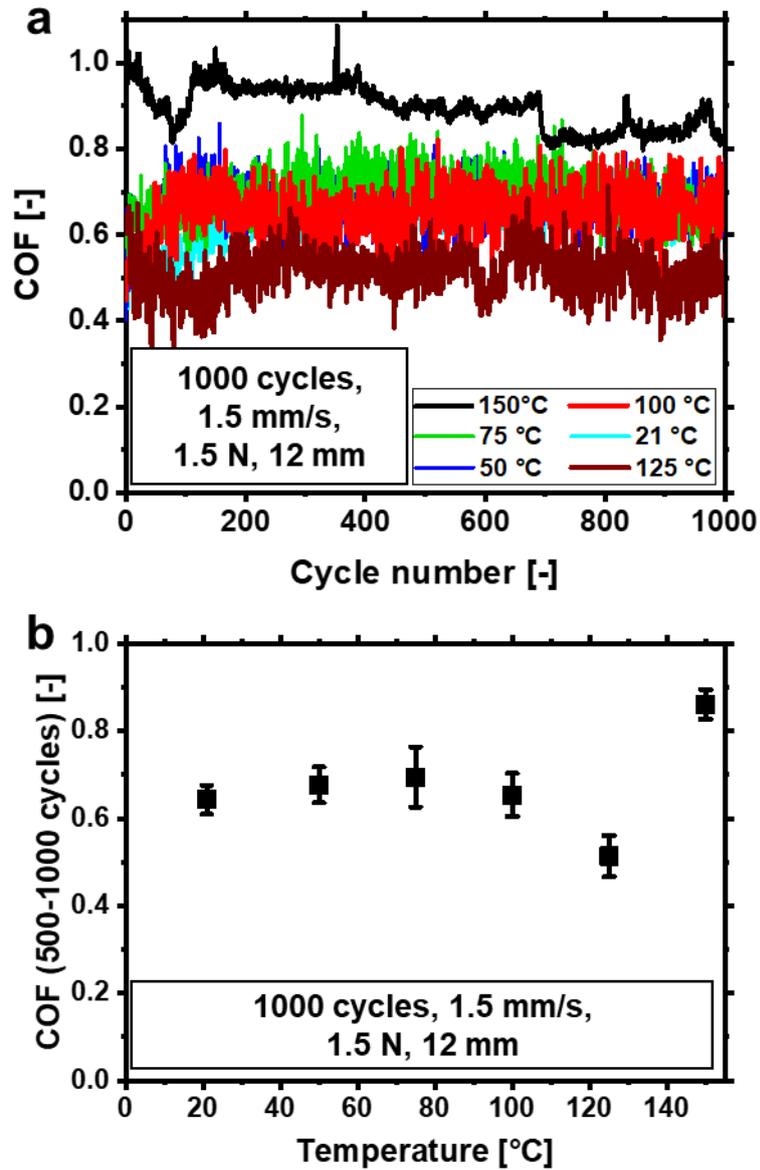

**Figure 2: Cross-sectional scanning electron microscopy images and maximum oxide thickness after 1,000 sliding cycles.** Images were taken parallel to sliding direction at (a) 21 °C, (b) 50 °C, (c) 75 °C, (d) 90 °C, (e) 100 °C, (f) 125 °C and (g) 150 °C sample temperature. (h) Average maximum oxide thickness with increasing temperature. Each data point represents the average of two experiments, except the experiments at 90 and 150 °C.

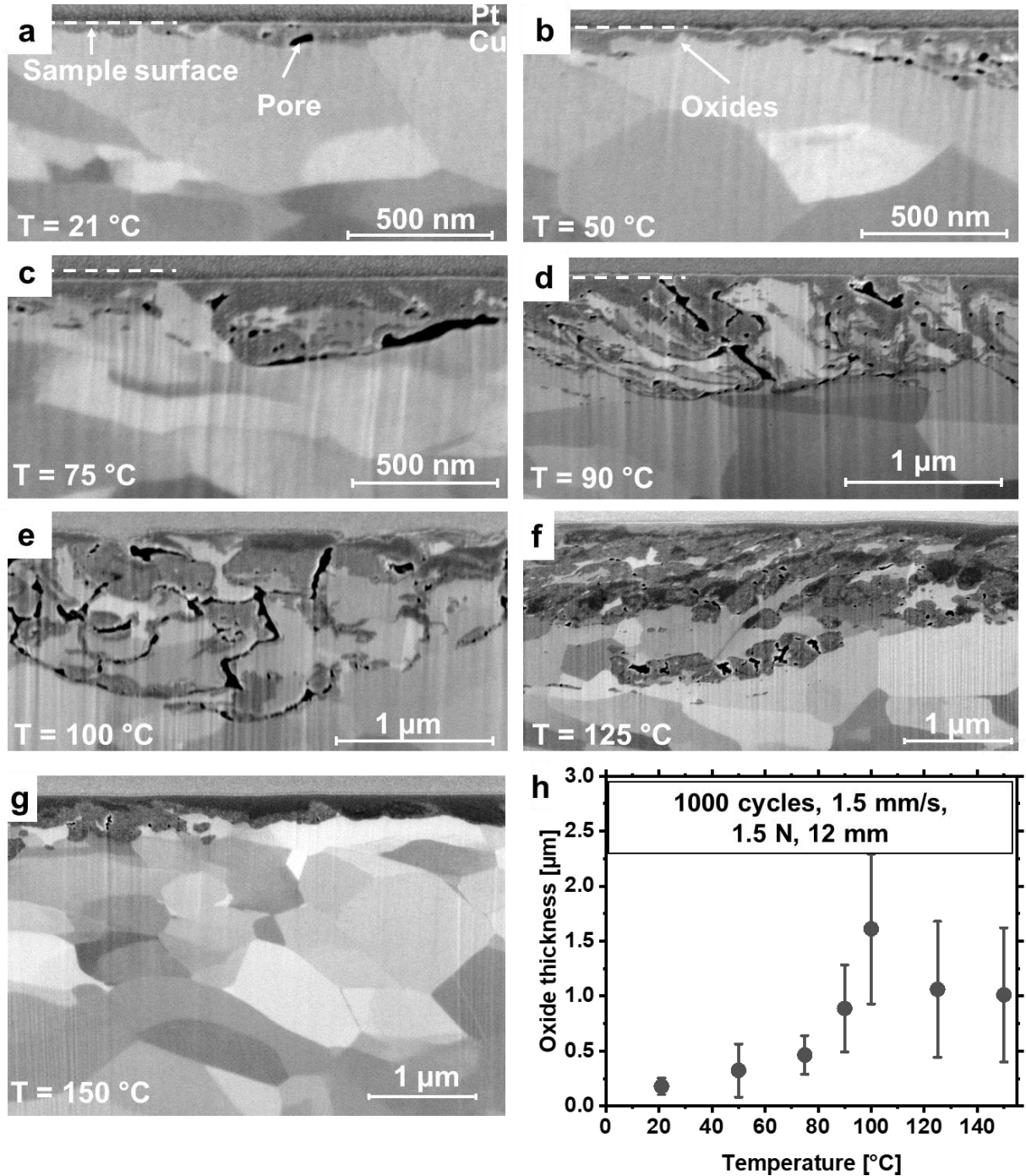

**Figure 3: Quantitative oxide and pore analysis.** (a+b) Average oxide and pore fraction over the material depth for the sample temperatures 21 °C (mean values of two experiments) and 150 °C (one experiment). The values are determined via image analysis from cross-sectional SE-SEM images and corrected for the 52° sample tilt. (c) Total oxide and pore area with increasing temperature. Triangles = oxide area, squares = pore area. Data points without standard deviation are results from single experiments. (d) CuO fraction with increasing sample depth z for 100 - 150 °C. For 150 °C, only one sample is evaluated. The experiments for 100 and 125 °C are averaged from two experiments.

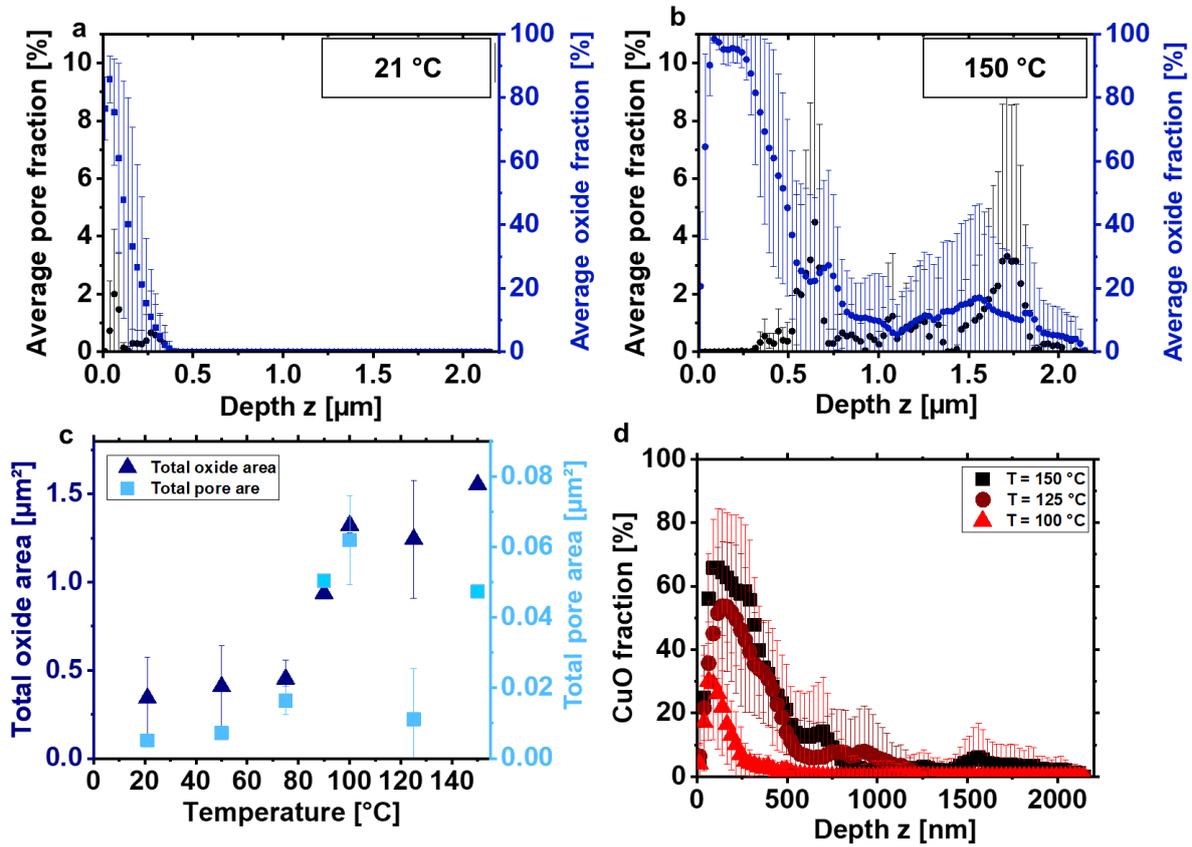

**Figure 4: Illustration of the quantitative oxide- and pore-fraction image analysis with ImageJ.** (front) Cross-sectional SE-SEM image of a sample after 1,000 cycles at 100 °C. (top) A sapphire sphere sliding against a copper plate: optical light microscopy image of the wear track, coloring of the surface through oxide formation while sliding. (right) Quantitative oxide and pore fraction analysis with ImageJ.

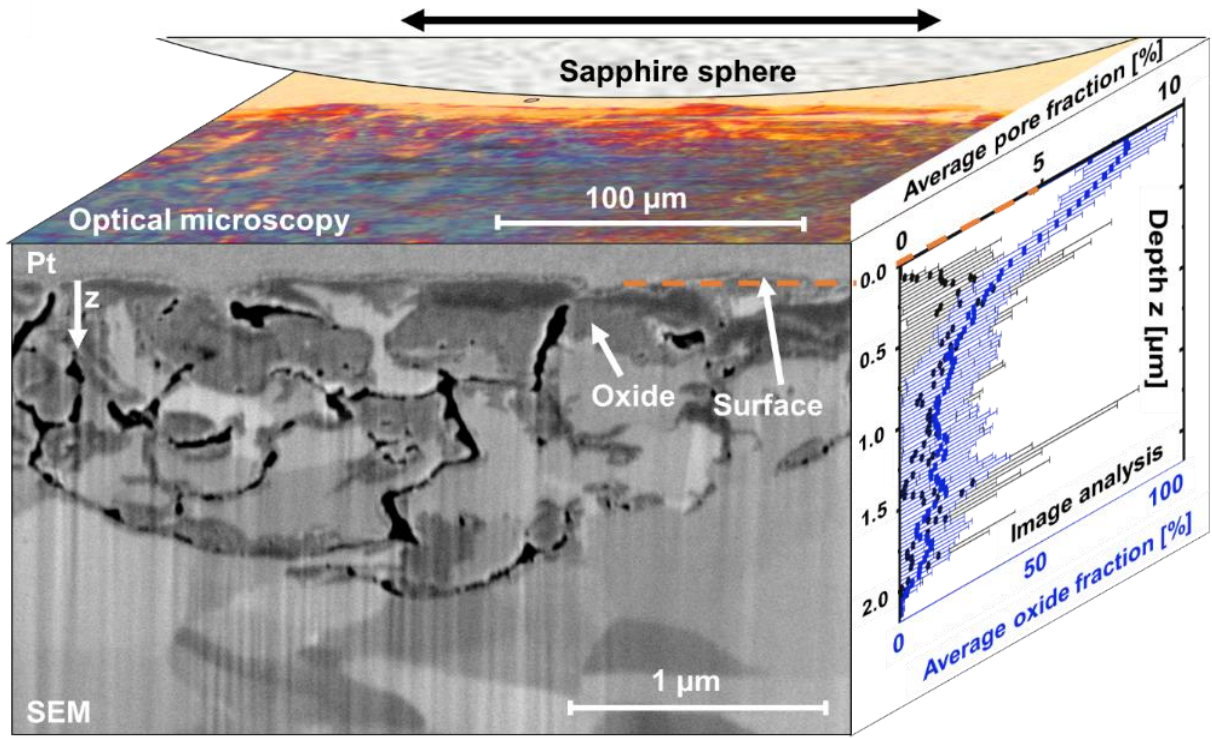
30

**Figure 5: Chemical composition of oxides after 1,000 cycles at 100 and 125 °C. Combined scanning transmission electron microscopy and energy-dispersive X-ray spectroscopy (STEM-EDS) measurements.** (a+d) Oxygen maps; (b+e) copper maps. (c+f) Element-concentration profile along the white lines in (b+e) for copper and oxygen, the lines indicate the composition for $Cu_2O$ and CuO.

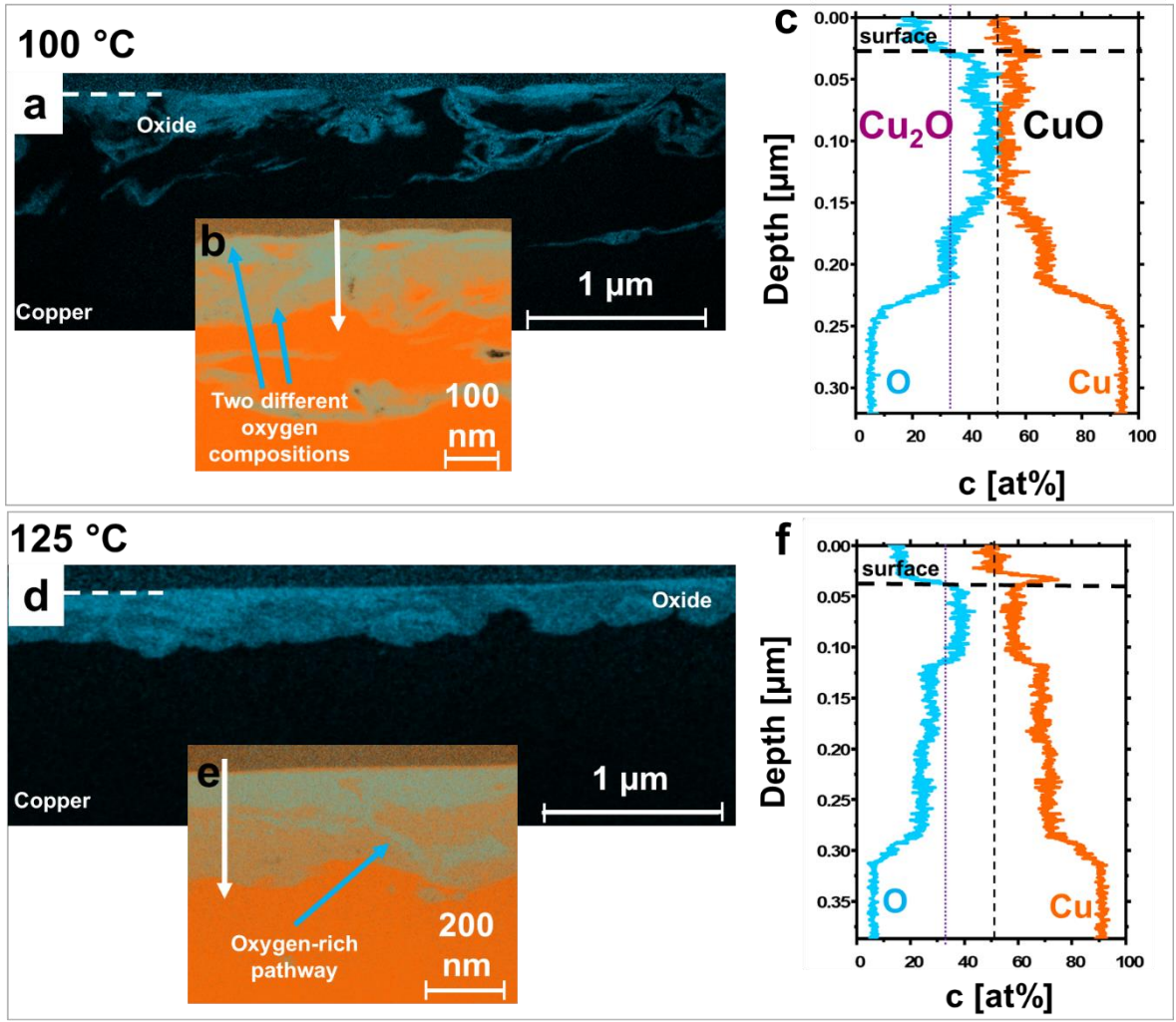

**Figure 6: TEM-EELS analysis after 1,000 cycles at 100 and 125 °C.** (a) STEM-BF-image and (b) O-K edge ELNES point analyses of marked areas in (a) for the 100 °C sample. (c) STEM-HAADF image and (d) EELS point analyses of the O-K- and Cu-L$_{23}$-edges of the marked areas in (c) for the 125 °C sample. The sample in (c) exhibits a layered structure with alternating CuO and Cu$_2$O-layers. A background substraction was performed for (b+d).

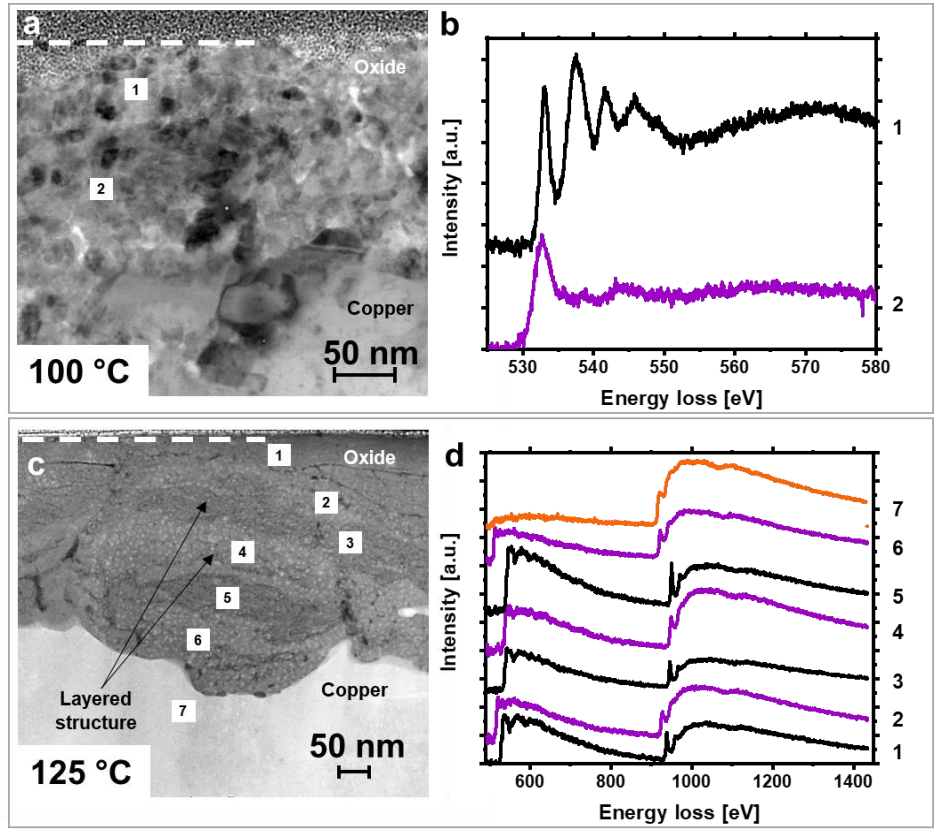

**Figure 7: Arrhenius-relation for the maximum effective diffusion coefficient of oxygen in copper over the inverse temperature.** Data points without standard deviations are single experiments. Values are calculated by use of the Einstein-relation.

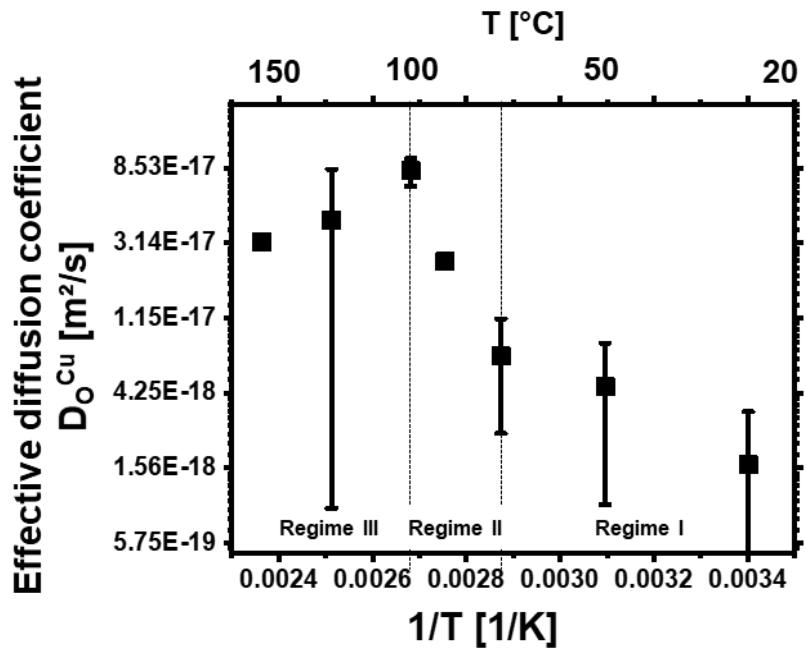

**Figure 8: Penetration profiles from quantitative oxide analysis.** (a) Schematic tracer distribution (orange for type B diffusion along a grain boundary after [38,41] ($d$ = grain size). (b) Schematic penetration profile for a bicrystal with type B diffusion after [41]. (c-f) Logarithmic oxide fraction vs. penetration depth $z^{6/5}$ for grain boundary diffusion (c+d) and $z$ for dislocation pipe diffusion (e+f). Average values, except single values for 90 and 150 °C. (g) Cross-sectional HAADF-STEM image after 1,000 cycles at 100 °C. The sample surface is marked by a dashed line. Each arrow marks the area where grain boundary(blue) or dislocation pipe (red) diffusion was dominating.

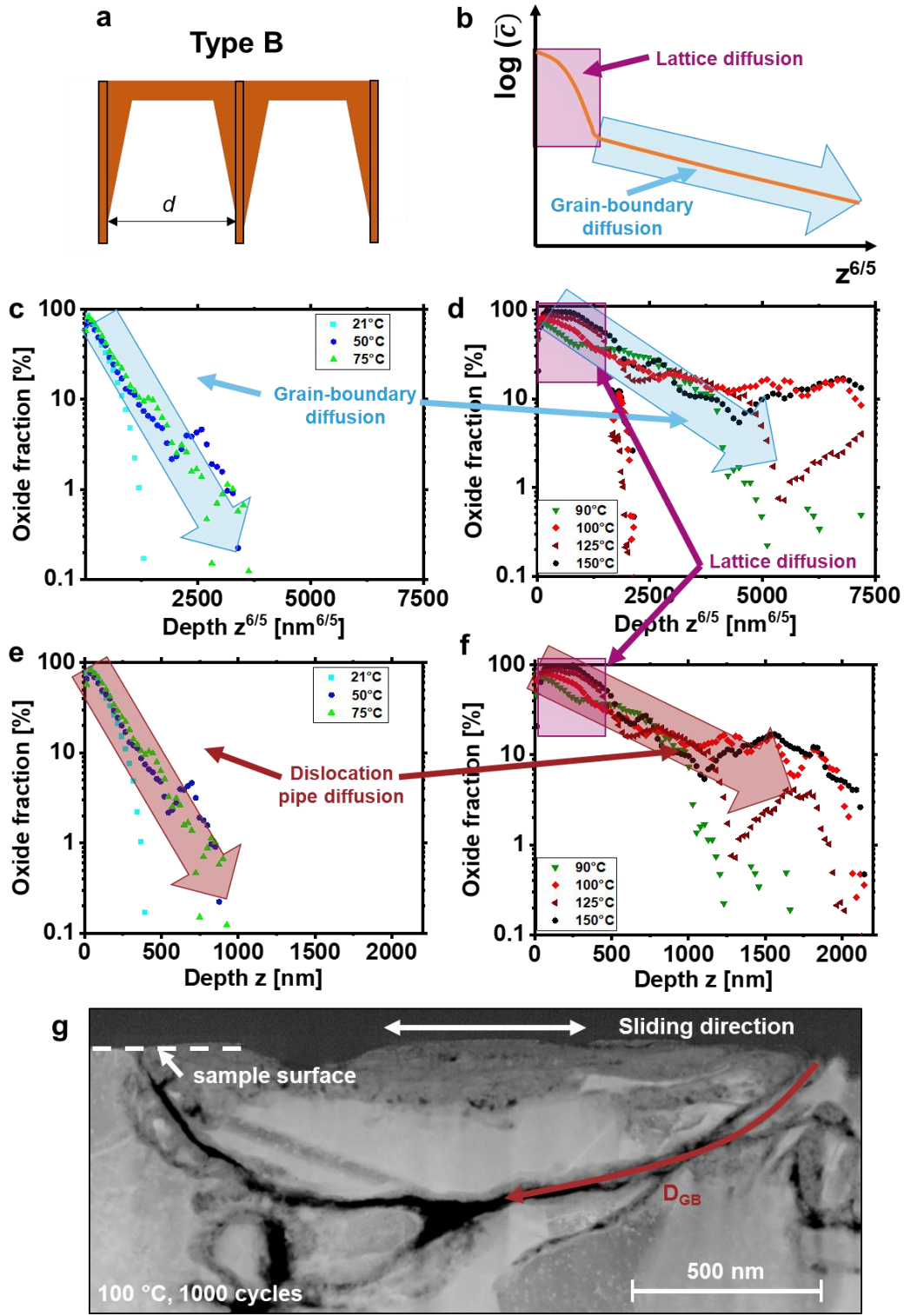